# Non-filamentary memristive switching in Pt/CuO$_x$/Si/Pt systems

L. L. Wei,[1,2] D. S. Shang,[2,3,5] J. R. Sun[2,5] S. B. Lee,[4] Z. G. Sun,[1] and B. G. Shen[2]

[1]State Key Laboratory of Advanced Technology for Material Synthesis and Processing, Wuhan University of Technology, Wuhan 430070, People's Republic of China

[2]Beijing National Laboratory for Condensed Matter Physics and Institute of Physics, Chinese Academic of Sciences, Beijing 100190, People's Republic of China

[3]I. Physikalisches Institut (IA), RWTH Aachen University, 52056 Aachen, Germany

[4]IBS Center for Functional Interfaces of Correlated Electron Systems & Department of Physics and Astronomy, Seoul National University, Seoul 151-747, Republic of Korea

E-mail: shangdashan@iphy.ac.cn and jrsun@iphy.ac.cn

**Abstract**

We report a memristive switching effect in the Pt/CuO$_x$/Si/Pt devices prepared by rf sputtering technique at room temperature. Different from other Cu-based switching systems, the devices show a non-filamentary switching effect. A gradual electroforming marked by resistance increasing and capacitance decreasing is observed in current-voltage and capacitance characteristics. By the Auger electron spectroscopy analysis, a model based on Cu ion and oxygen vacancy drift, and thickness change of the SiO$_x$ layer at the CuO$_x$/Si interface was proposed for the memristive switching and gradual electroforming, respectively. The present work would be meaningful for the preparation of forming-free and homogeneous memristive devices.

1. Introduction

Nanoionics-based memristive devices (NIMD) have attracted intensive attention in recent years due to their potential application in resistive random access memory (RRAM) as well as reconfigurable logic devices [1,2]. The basic structure of NIMD consists of an electrode/electrolyte/electrode sandwich structure. According to the charge type of the active ions in switching, NIMD can be classified into two categories: cation-based device and anion-based device. The typical active cations are $Cu^{2+}$ and $Ag^+$, and the familiar anion is $O^{2-}$. The transport of $O^{2-}$ is equivalent to that of oxygen vacancy. Many kinds of materials can be used as electrolyte, such as GeSe [3], CuS [4], $SiO_2$ [5], $TaO_x$ [6], with metal ions and $TiO_{2-x}$ [7], $SrTiO_{3-x}$ [8], Cr-$SrTiO_{3-x}$ [9], $Pr_{0.7}Ca_{0.3}MnO_{3-x}$ [10], with oxygen vacancies. For the cation-based NIMD, two processes, ion drift and redox at anode/cathode, generally occur under external electric field, leading to metal filaments in electrolyte [1-6,11]. The memristive switching is realized by the connection and rupture of the metal filament. Although the detailed processes are still under debate [11,12], they are believed to be stochastic in nature, leading to the non-uniformity of switching parameters [2]. Moreover, the scaling potential of the NIMD suffers from the tiny size of the filament, which will lead to a large current density in the switching process as well as for the selector devices [13]. Due to these drawbacks, NIMD devices without uncontrollable filaments is desirable for super integrated devices. In this paper, we present an investigation on Cu-based NIMD with the structure of Pt/CuO$_x$/Si/Pt. This device displays a special electroforming accompanied by resistance increasing and capacitance reducing, and a non-filamentary memristive switching. Based on the analyses of valence state of each element around interface after repeated resistance switching, the physical mechanism of the special electroforming and memristive switching is discussed.

2. Experiment

Samples were prepared by the technique of radio-frequency magnetron sputtering at ambient temperature. A Si layer of 40 nm in thickness was first deposited on a commercial Pt/Ti/SiO$_2$/Si substrate in an Ar atmosphere of 1 mTorr. Then a CuO$_x$ layer of 25 nm was deposited in the Ar/O$_2$ mixed gas atmosphere of 5 mTorr with Ar:O = 5:1. As top electrode, a Pt layer of 30 nm was finally deposited in the Ar atmosphere of 5 mTorr. The Pt/CuO$_x$ layers were patterned, by the photolithography and lift-off technique, to square arrays with the sizes ranging from 75×75 μm$^2$ to 175×175 μm$^2$. The as-deposited Si and CuO$_x$ layers are determined to be amorphous since no peaks can be observed in the X-ray diffraction spectra (see figures S1 and S2 in supplementary data). For comparison, the devices with a stack structure of Pt/CuO$_x$(25nm)/Pt and Pt/Si(40nm)/Pt were also prepared under the same conditions. The current-voltage (*I-V*) characteristics were measured by a Keithley 2601 SourceMeter. Electric polarity directing from top to bottom electrodes was defined as positive. Impedance spectroscopy in the frequency range of 40 Hz–110 MHz was measured by an Agilent 4294A Precision Impedance Analyzer under an ac bias of 0.4 V. The depth profile and the chemical compositions of the CuO$_x$/Si interfaces were analyzed by an Auger electron spectrometer at a sputtering rate of 4.4 nm/min. All the measurements present in this paper were taken on the devices with the electrodes size of 125×125 μm$^2$ and all the device resistances were read under a dc bias of 0.4 V except for special specification.

## 3. Results and discussion

Figure 1(a) shows the *I-V* characteristics of the Pt/CuO$_x$/Si/Pt device, obtained by cycling dc bias along 0 V→+4 V→0 V→−4 V→0 V at a ramping speed of 0.2 V/s. A current compliance ($I_{com}$) of 10 mA was selected to avoid permanent breakdown. The *I-V* curves exhibit a pronounced electric hysteresis, which is a signature of memristive switching. The device transited from a high resistance state (HRS) to a low resistance state (LRS) with the increase of positive bias and switched back after the sweeping of negative bias. The high to low resistance ratio, recorded under a voltage bias of 0.4 V, is larger than 10$^2$ even after 100 cycles (inset in figure 1(c)). Figure 1(b) shows the typical retention behavior of the device under the reading voltage of 0.4 V. No significant changes were observed up to 10$^2$ seconds for the HRS, whereas the LRS resistance exhibited a slow increase with time.

In general, a so-called electroforming process is needed for filament-type memristive switching. In this process, the device shows an abrupt switching from the initial resistance state (IRS) to a low resistance state, and this switching needs an electric-field higher than that for following switchings [3-6,13]. However, this type of electroforming did not appear in the Pt/CuO$_x$/Si/Pt devices. As shown in figure 1(a), the device exhibited a smooth *I-V* relation, especially for the first few electric cycles, and no obvious abrupt resistance switching was observed. Meanwhile, the *I-V* loops underwent a gradual downward shift with electric cycling, which is an indication of progressive resistance growth, and stable *I-V* loops were obtained only after about 30 electric cycles. It means that the forming process which leads to the electric structure favoring steady resistance switching lasts for about 30 cycles. To distinguish from the previous one, we call this forming process as gradual electroforming (GE).

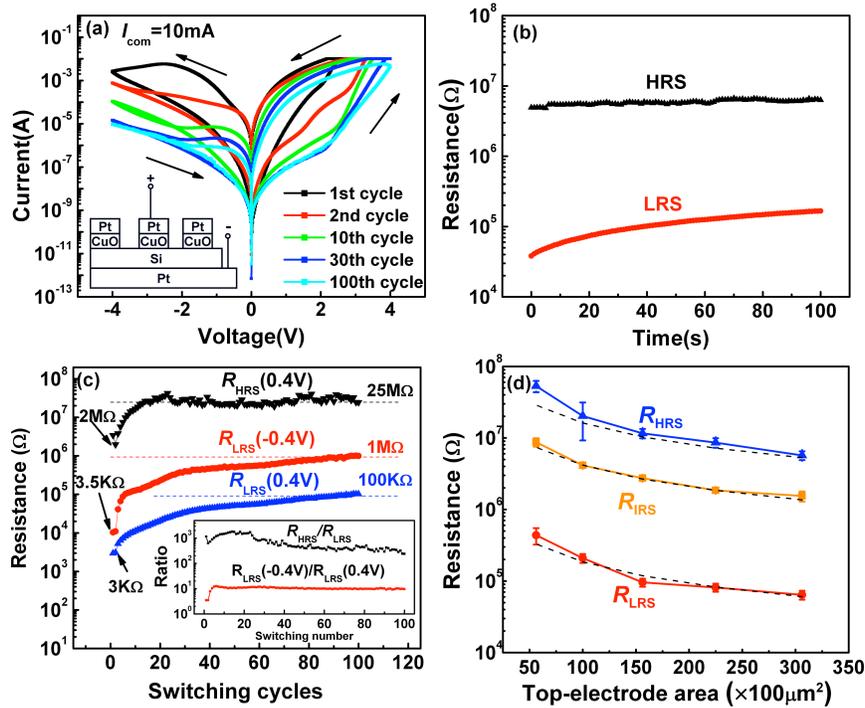

**Figure 1.** (a) *I-V* characteristics of the Pt/CuO$_x$/Si/Pt devices with a current compliance of 10 mA. Inset: Schematic of the device and measurement. (b) Retention behavior of the HRS and LRS under a reading voltage of 0.4 V. (c) Variation of the resistance values in the HRS ($R_{HRS}$) and LRS ($R_{LRS}$) as a function of the switching cycles. Inset: Variation of the $R_{HRS}/R_{LRS}$ and rectifying ratio in the LRS. (d) Resistance variation as a function of the electrode area. The resistances were read from five electrode points for each electrode area in the same devices with reading voltage of 0.4 V after the GE process. Dashed curves are the reference for the resistance inversely proportional to the electrode area.

The GE assigns the device three electric behaviors different from those yielded by the electroforming that generates a conduction filament. First, resistances of both the HRS and LRS increased with *I-V* cycling (figure 1(c)). As a consequence, the resistance of the HRS is larger than that of the IRS. Second, the *I-V* characteristics in the LRS evolved from symmetry to asymmetry after repeated electric cycles. As shown in the inset of figure 1(c), the ratio of the resistances read under 0.4 V and -0.4 V, respectively, grew from ~1.2 to ~10 after the first 5 cycles and then became stable. Third, resistances of both the HRS and LRS after the GE showed a strong dependence on electrode area (figure 1(d)). It is an indication of non-filament-type switching.

To examine which component of the device is responsible for the memristive switching, the *I-V* characteristics of the Pt/CuO$_x$/Pt and Pt/Si/Pt devices were also studied. As shown in figure 2, both the Pt/CuO$_x$/Pt and the Pt/Si/Pt devices exhibit linear-like *I-V* relations, indicating the Ohmic character of the Pt/CuO$_x$ and Si/Pt contacts. The device resistances are ~51 Ω and ~36 Ω for the Pt/CuO$_x$/Pt and Pt/Si/Pt devices, respectively, significantly lower than that of the Pt/CuO$_x$/Si/Pt devices (~1 MΩ). CuO$_x$ is a semiconductor and the high conductance could be ascribed to oxygen vacancies. Usually amorphous Si has a high resistivity of ~10$^5$ Ω·cm, corresponding to about 1 kΩ for our devices. The lower resistance may originate from structure defects. The Si layer is actually composed of relatively incompact nanometer-sized (~70 nm) grains (see figure S3 in supplementary data). By comparing the three devices, we came to the conclusion that the high resistance of Pt/CuO$_x$/Si/Pt arose from the CuO$_x$/Si interface, and it is this interface that is responsible for the memristive effect.

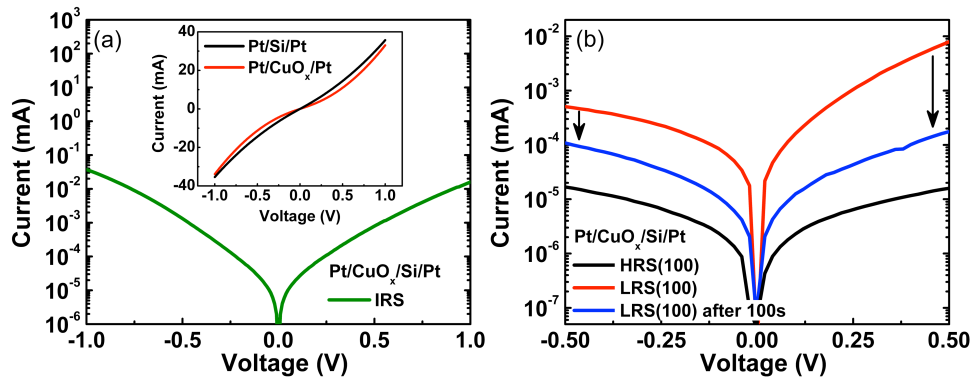

**Figure 2.** (a) *I-V* curves of the Pt/CuO$_x$/Pt, Pt/Si/Pt, and Pt/CuO$_x$/Si/Pt devices in the IRS. (b) *I-V* curves of the Pt/CuO$_x$/Si/Pt devices in the HRS and LRS behind 100 switching cycles, and LRS after 100 s behind 100 switching cycles.

Besides resistance switching, a capacitance switching was also observed in the Pt/CuO$_x$/Si/Pt device. Figure 3 shows the typical impedance spectra of the IRS, HRS, and LRS. The impedance spectra can be well fitted by an equivalent circuit consisted of one paralleled $R$ and $C$ in series with a resistor ($R_0$). $R_0$ is lower than 40 Ω, representing the resistance from electrode contacts, leading wires, and the body of the device, while the paralleled $R$ and $C$ mimic the CuO$_x$/Si interface. Capacitances obtained by curve-fitting are 210 pF, 140 pF, and 170 pF for the IRS, HRS, and LRS, respectively. The capacitance of the LRS was always larger than that of the HRS, which could be an indication for the reduction of effective thickness of insulating layer at the CuO/Si interface. Corresponding to the GE, as illustrated in figure 4(a), the capacitance in either HRS ($C_{HRS}$) or LRS ($C_{LRS}$) and the $C_{HRS}/C_{LRS}$ ratio all showed a gradual decrease with electric cycling before stabilized after 30 cycles.

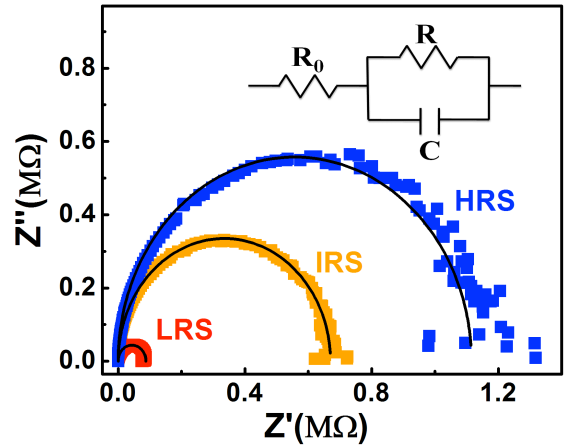

**Figure 3.** Cole-Cole plots of the Pt/CuO/Si/Pt device in the IRS, HRS, and LRS. Inset: The equivalent circuit for curve fitting. The data used for fitting are $R_0$=18 Ω, $R_{HRS}$=1.11 MΩ, $C_{HRS}$=140 pF, $R_{LRS}$=87.8 kΩ, $C_{LRS}$=170 pF, $R_{IRS}$=670 kΩ, and $C_{IRS}$=210 pF.

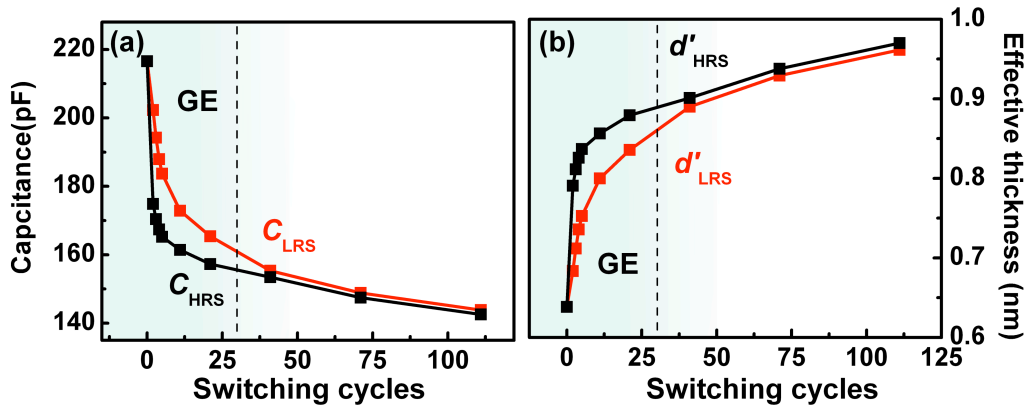

**Figure 4.** Variation of (a) the capacitance of the Pt/CuO$_x$/Si/Pt devices and (b) the calculated effective thickness as a function of switching cycles.

There are possible reasons for the high resistance of the CuO$_x$/Si interface as well as the electric cycle-induced resistance/capacitance change. The first one is the presence of an interfacial barrier. As well established, interfacial barrier favored rectifying *I-V* characteristics. However, the *I-V* curves in the HRS and IRS are symmetric (figure 2), where the measurement is performed within bias range between -1 V and +1 V to avoid resistance switching. Although the *I-V* curve in the LRS is asymmetric, it became symmetric at 100 s after switched to LRS (figure 2(b)). This implies that the rectification shown in figure 1(a) is a metastable behavior rather than an effect due to interfacial barrier. These results excluded the existence of the barriers. Another reason is the formation of an insulating interfacial layer. According to the formation energies of CuO (-129.7 kJ/mol), Cu$_2$O (-146 kJ/mod), SiO (-342 kJ/mol), Si$_2$O (-426 kJ/mol), and SiO$_2$ (-911 kJ/mol) [14,15], the following chemical reactions could take place spontaneously around the CuO$_x$/Si interface:

$$2CuO_x + Si \rightarrow Cu_2O + SiO_x; \quad CuO_x + Si \rightarrow Cu + SiO_x \tag{1}$$

It is therefore possible to form a transitional layer containing SiO$_x$, Cu, and Cu$_2$O at the CuO$_x$/Si interface in the process of sample preparation and the following electric cycling.

In order to verify this conclusion, the composition and chemical states of the CuO$_x$/Si interface were investigated by AES depth analysis. Figures 5(a) and (b) show the AES depth profile spectra of the CuO$_x$/Si interfaces in the IRS and HRS(100), where HRS(100) is the HRS(100) after $10^2$ 0→4→-4→0 V cycles. As expected, a transitional layer consisting of Cu, Si, and O was detected at the CuO$_x$/Si interface. The valence states of Cu, and O at various depths can be determined based an analysis of the line shape in AES spectra. As shown in figures 5(c)-(f), the kinetic energy of Cu *LMM* decreased from 915 eV (Cu$^{2+}$) to 914 eV (Cu$^+$) and then to 916 eV (Cu$^0$) when the depth extended from the CuO$_x$ layer to the CuO$_x$/Si interface in both the IRS and HRS(100), indicating a progressive reduction of Cu$^{2+}$ to Cu$^+$ and then Cu$^0$ [16]. Correspondingly, the kinetic energy of O *KLL* decreased from 511 eV (CuO) to 506 eV (SiO$_x$) [17]. These results are consistent with the chemical reactions described by Eq. (1), and demonstrate the emergence of a SiO$_x$ layer around the CuO$_x$/Si interface, coexisted with Cu$_2$O and Cu.

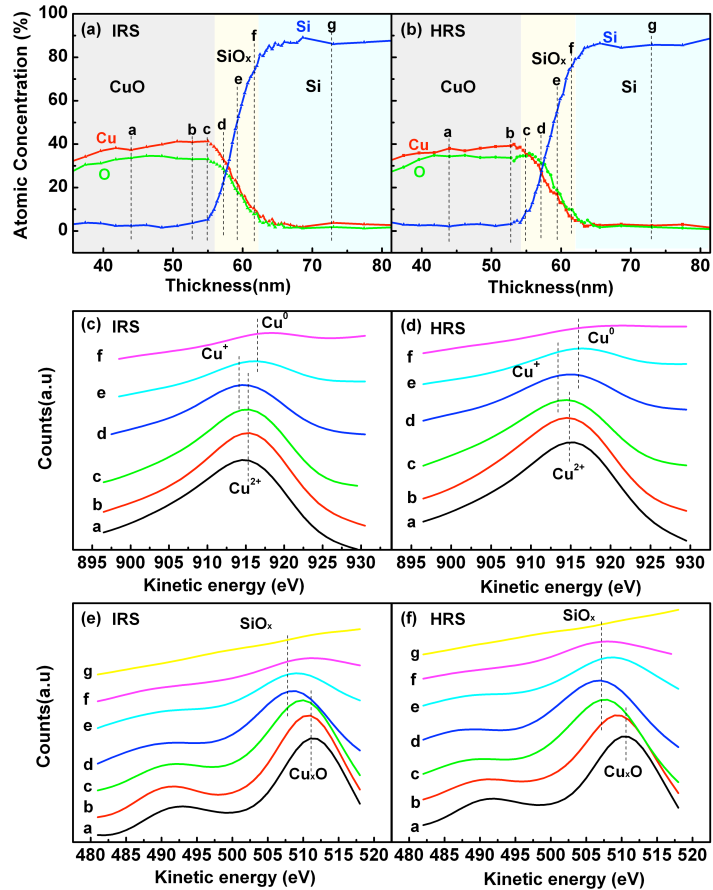

**Figure 5.** (a) and (b) show the AES depth profile spectra at the CuO$_x$/Si interface in the IRS and HRS(100), respectively. (c) and (d) show the AES line shape analysis of Cu *LMM* in the IRS and HRS(100), respectively. (e) and (f) show O *KLL* in the IRS and HRS(100), respectively. Lines from a to g identify the different depths.

The AES spectra also revealed a thicknesses change of the SiO$_x$ layer from ~5 nm to ~7 nm when the resistive state of the device switched from IRS to HRS(100). We also measured the AES spectra of the LRS(100). However, no difference between the HRS(100) and LRS(100) can be distinguished, which indicates the stabilization of the interfacial layer after 100 electric cycles. These results are consistent with the capacitance decrease with electric cycling. The capacitance is closely related to the SiO$_x$ layer thickness according to the formula $d'=d/\varepsilon_r=\varepsilon_0 S/C$, where $d$ is the layer thickness, $\varepsilon_r$ is the relative permittivity, $\varepsilon_0$ is the vacuum permittivity, and $S$ is the electrode area. As shown in figure 4(b), the effective thickness $d'$ first grew from ~0.68 nm to ~0.89 nm in the LRS and from ~0.79 to ~0.90 nm in the HRS during the first ~30 cycles corresponding to the GE. With further cycling, it became relatively stable and its difference in the LRS and HRS diminished progressively, and was almost negligible above ~50 cycles.

Mechanism for the memristive switching in the Cu/electrolyte systems with the electrolytes of SiO$_2$ [5], Ta$_2$O$_5$ [6], GeO$_x$ [18], ZnO [19], and HfO$_2$ [20], has been reported, and the formation and rupture of Cu filament that link the top and bottom electrodes were believed to be responsible for the resistance change. However, this model cannot explain the GE, the capacitance switching, and the non-filamentary memristive switching effects occurred in the Pt/CuO$_x$/Si/Pt devices. According to the experimental results, a model based on the drift of Cu$^{z+}$ and the oxidation of Si at the CuO$_x$/Si interface is proposed. A SiO$_x$ layer was formed at the CuO$_x$/Si interface in the process of sample preparation, resulting in a high interfacial resistance of the IRS (figure 6(a)). The positive biases drove the Cu$^{z+}$ ions in the CuO$_x$ layer to drift through the SiO$_x$ layer and accumulate at the Si layer surface, due to the lower mobility of Cu$^{z+}$ in Si than in SiO$_x$ [21]. At the same time, oxygen vacancies could also be driven into the SiO$_x$ layer. Both the Cu$^{z+}$ ions and oxygen vacancies act as trapping states, providing paths for electron tunneling through the SiO$_x$ layer. Different from the previously studied Cu/electrolyte devices, the Cu$^{z+}$ ions in our samples were provided by the CuO$_x$ layer, rather than the anodic dissolution of a Cu electrode, and Cu$^+$ and Cu$^0$ were already existed in the SiO$_x$ layer even before the forming operation according to the AES data. As a result, plenty of paths for electron tunneling could be formed simultaneously in the SiO$_x$ layer with an addition even minor Cu$^{z+}$ or oxygen vacancies, yielding the resistance of electrode-area dependence.

Under negative biases, the Cu$^{z+}$ moved back to the CuO$_x$ layer and, in the meantime, the oxygen vacancies migrate out of the SiO$_x$ layer. The latter, which is equivalent to introducing O$^{2-}$ into SiO$_x$, will cause a further oxidation of SiO$_x$ thus a growth in the SiO$_x$ layer thickness (figure 6(c)). Consequently, the HRS showed a higher resistance than the IRS. When applying next positive bias, Cu$^{z+}$ and oxygen vacancies drifted into the SiO$_x$ layer again. However, it may be difficult to totally extract the previously incorporated O$^{2-}$ from the SiO$_x$ layer due to the strong Si-O bonding as well as the low formation energy of SiO$_x$. This means that the SiO$_x$ layer will become thicker than that of the last LRS (figure 6(d)). Due to the similar reason in case of the last negative biases, when the device was driven to the HRS again, the SiO$_x$ layer became even thicker. This actually implies a rapid growing in the SiO$_x$ layer thickness, thus in resistance, with electric cycling in the GE process (figure 1(c)). After the GE process, a relatively stable SiO$_x$ layer was formed, and further electric cycling only caused minor changes in layer thickness. These analyses explain the detected capacitance thus the effective layer thickness change shown in figure 4.

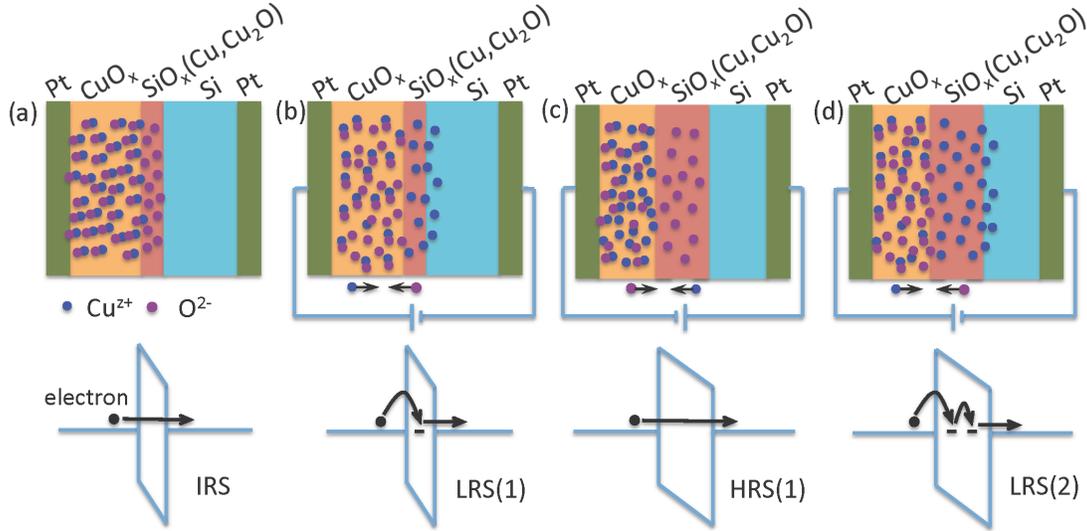

**Figure 6.** Schematics of the proposed chemical (top) and electronic (bottom) modifications between the (a) IRS, (b) the first time of LRS, (c) the first time of HRS, and (d) the second time of LRS of the memristive switching in the Pt/CuO$_x$/Si/Pt devices. For simplicity, the oxygen vacancies is expressed by the oxygen ions (O$^{2-}$).

As has been reported, shapes of the metallic filaments are mainly determined by the field-induced crystallization process [1,11]. In this process cations are reduced to metal atoms and then precipitate. The precipitation usually occurs when the dissolved ions are so supersaturated that reach a threshold concentration to initiate the nucleation/growth of the precipitates [22]. Due to the large lattice defects as well as large specific surface of the incompact amorphous Si layer, the solubility of Cu in the Si layer could relatively high. In this case, the amorphous Si layer may act as an ion sink, suppressing the Cu precipitation and then the filament-type growth of the precipitates. Therefore, the memristive switching shows a non-filamentary feature. As for the relaxation of the LRS with time (figure 1(b)), it can be attributed to a self-diffusion of Cu$^{z+}$ out of SiO$_x$ due to its high concentration in SiO$_x$. The ion diffusion-induced by concentration gradient could also be used to explain the metastable rectifying behavior observed in the LRS (figure 1(a)). Since the Cu$^{z+}$ diffusion would be enhanced under negative bias, it accelerated the device resistance increase, inducing a rectifying behavior in the LRS. After waiting for 100 s after switched to the LRS, the Cu$^{z+}$ self-diffusion was almost completed, and the *I-V* curve became stable and non-rectifying (figure 2). From the point of view of memory devices, the GE is still undesired as well the instable LRS. Improving these properties could be expected by, for example, preparing an inert layer at the CuO$_x$/Si interface where the further oxidation can be avoided and designing proper ion diffusion barrier to restrain the ion self-diffusion. Finally, it should be pointed out that the non-filamentary switching was only demonstrated within certain range of electrode sizes in this work. Thus, further studies, such as scaling the device size down to nanoscale, are still required.

## 4. Conclusions

In conclusion, we have investigated the memristive switching characteristics of the Pt/CuO$_x$/Si/Pt devices based on the *I-V* curves and capacitance analyses. The devices showed a GE process and a non-filamentary switching behavior. The device resistances on both the HRS and LRS increased with *I-V* cycling as well as the device capacitances decreased gradually before reached stabilization after GE process. AES depth analysis revealed the emergence of a SiO$_x$ layer around the CuO$_x$/Si interface, coexisted with Cu$_2$O and Cu, and the thickness growth of this layer under external field. Based on the experimental results, we propose that the Cu$^{z+}$ and oxygen drift into/out of the SiO$_x$ layer driven by the electric-field is responsible for the memristive switching, and the GE process is attributed to the thickness increase of the SiO$_x$ layer due to the further oxidation under the negative bias condition.

## Acknowledgements

This work has been supported by the National Basic Research of China, the National Natural Science Foundation of China, the Knowledge Innovation Project of the CAS, the Alexander von Humboldt Foundation (for D.S.S) and the Institute of Basic Science (for S.B.L, Grant No. 2012-0005847) funded by the Korea government (MEST).

Supplementary Data

# Non-filamentary memristive switching in Pt/CuO$_x$/Si/Pt systems

L. L. Wei,[1,2] D. S. Shang,[2,3,5] J. R. Sun,[2,5] S. B. Lee,[4] Z. G. Sun,[1] and B. G. Shen[2]

[1]State Key Laboratory of Advanced Technology for Material Synthesis and Processing, Wuhan University of Technology, Wuhan 430070, People's Republic of China

[2]Beijing National Laboratory for Condensed Matter Physics and Institute of Physics, Chinese Academic of Sciences, Beijing 100190, People's Republic of China

[3]I. Physikalisches Institut (IA), RWTH Aachen University, 52056 Aachen, Germany

[4]IBS Center for Functional Interfaces of Correlated Electron Systems & Department of Physics and Astronomy, Seoul National University, Seoul 151-747, Republic of Korea

[5]E-mail: shangdashan@iphy.ac.cn and jrsun@iphy.ac.cn

**Supporting Information:**

1) X-ray diffraction (XRD) pattern of the CuO films on Si/Pt/Ti/SiO$_2$/Si substrate.

2) XRD pattern of the Si films on single crystal Al$_2$O$_3$ substrate.

3) Atomic force microscopy (AFM) of the Si films on Pt/Ti/SiO$_2$/Si substrate

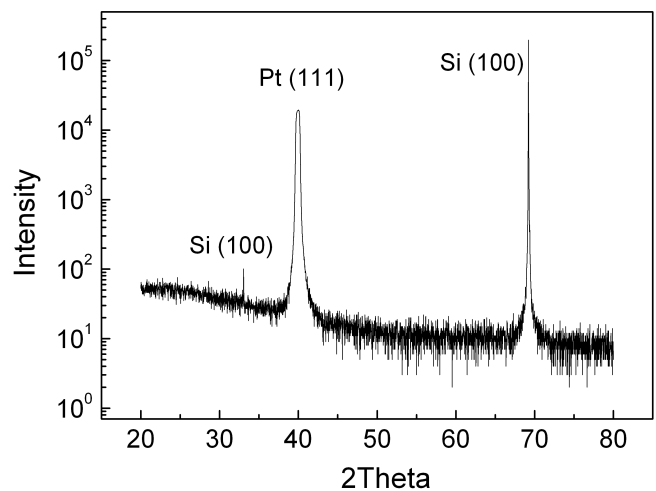

**Figure S1.** XRD pattern of the CuO films on Si/Pt/Ti/SiO$_2$/Si substrate, showing the structure of the CuO films is amorphous. The CuO films were prepared by radio-frequency magnetron sputtering at ambient temperature in the Ar/O$_2$ mixed gas atmosphere of 5 mTorr with Ar:O = 5:1.

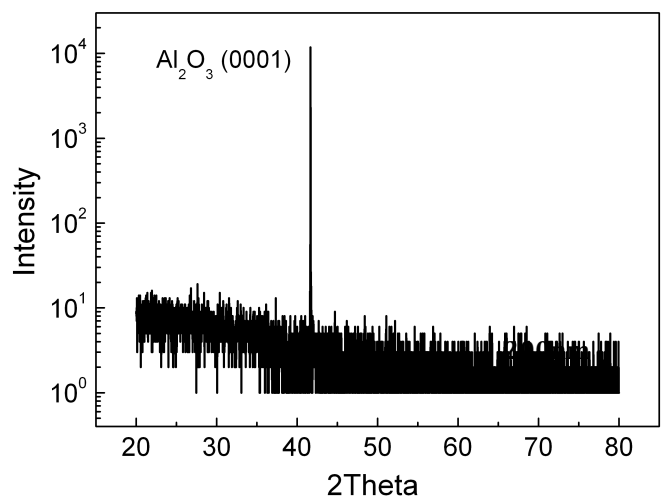

**Figure S2.** XRD pattern of the Si films on single crystal Al$_2$O$_3$ substrate, showing the structure of the Si films is amorphous. The Si films were prepared by radio-frequency magnetron sputtering at ambient temperature in the Ar atmosphere of 1 mTorr.

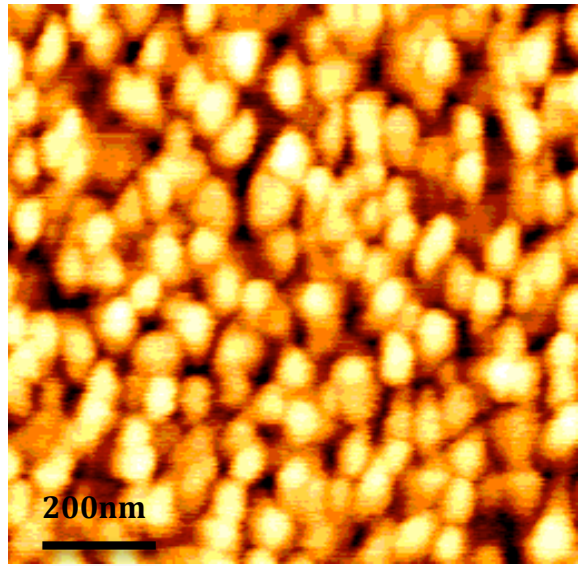

**Figure S3.** AFM topography of the Si films on Pt/Ti/SiO$_2$/Si substrate. The as-deposited Si layer is actually composed of relatively incompact nanometer-sized (~70 nm) grains